\documentclass[aps,prb,twocolumn,floatfix,superscriptaddress]{revtex4}
\usepackage{graphicx}
\usepackage{natbib}
\usepackage{epstopdf}
\usepackage{amsmath}
\usepackage{amssymb}
\usepackage{mathbbol}
\usepackage{bm}
\usepackage{upgreek}
\usepackage{xcolor}
\usepackage[normalem]{ulem}
\DeclareMathOperator{\Tr}{Tr}

\newcommand{\e}{\mathrm{e}}
\newcommand{\abs}[1]{\left\vert#1\right\vert}
\newcommand{\beq}{\begin{equation}}
\newcommand{\eeq}{\end{equation}}
\newcommand{\beqy}{\begin{eqnarray}}
\newcommand{\eeqy}{\end{eqnarray}}
\newcommand{\inc}{\includegraphics}
\newcommand{\bit}{\begin{itemize}}
\newcommand{\eit}{\end{itemize}}
\newcommand{\bmat}{\begin{pmatrix}}
\newcommand{\emat}{\end{pmatrix}}

\begin{document}

% title, authors etc.

\title{Coherently driven microcavity-polaritons and the question of superfluidity}
\author{R. T. Juggins}
\email{richard.juggins@gmail.com}
\affiliation{Department of Physics and Astronomy, University College London,
Gower Street, London, WC1E 6BT, United Kingdom}
\author{J. Keeling}
\affiliation{SUPA, School of Physics and Astronomy, University of St Andrews, KY16 9SS, United Kingdom} 
\author{M. H. Szyma\'nska}
\email{m.szymanska@ucl.ac.uk}
\affiliation{Department of Physics and Astronomy, University College London,
Gower Street, London, WC1E 6BT, United Kingdom}
\date{\today}

\begin{abstract}

\noindent Due to their driven-dissipative nature, photonic quantum fluids present new challenges in understanding superfluidity. Some associated effects have been observed, and notably the report of nearly dissipationless flow for coherently driven microcavity-polaritons was taken as a smoking gun for superflow. Here we show that the superfluid response --- the difference between responses to longitudinal and transverse forces --- is zero for coherently driven polaritons. This is a consequence of the gapped excitation spectrum caused by external phase locking. Furthermore, while a normal component exists at finite pump momentum, the remainder forms a rigid state that is unresponsive to either longitudinal or transverse perturbations.  Interestingly, the total response almost vanishes when the real part of the excitation spectrum has a linear dispersion, which was the regime investigated experimentally. This suggests that the observed suppression of scattering should be interpreted as a sign of this new rigid state and not a superfluid.
\clearpage
\end{abstract}

\maketitle

One of the most spectacular emergent effects in quantum physics is that of superfluidity, which was first observed as a set of peculiar flow properties in liquid helium when cooled below 2.17K \cite{leggett1999superfluidity}. Understanding collective effects such as dissipationless flow, lack of response to transverse perturbations,
quantised vortices, and metastable persistent currents, has been the aim of much theoretical work \cite{leggett2006quantum, pitaevskii2003bose, baym1968microscopic, nozieres1999theory, griffin1993excitations}, particularly with respect to systems in thermodynamic equilibrium. Extending these well-established ideas to driven-dissipative systems, which do not thermalise due to constant pumping and decay, has however proved contentious \cite{lagoudakis2008quantized,wouters2010superfluidityPRL,keeling2011superfluid,janot2013superfluid,gladilin2016normal,wachtel2016electrodynamic,keeling2017superfluidity}.

While dissipationless flow, explained by the Landau criterion \cite{landau1941theory}, is perhaps the most famous property of superfluids, arguably the most fundamental is that they do not respond to transverse perturbations \cite{nozieres1999theory}: that is, the bulk of the fluid is irrotational. 
Crucially, this difference between longitudinal and transverse response onsets sharply at the phase transition, wheras perfectly dissipationless flow only occurs at absolute zero temperature, where the normal component vanishes.  Thus, to clearly distinguish a true superfluid from a fluid merely with low viscosity, one needs to focus on the response functions. This difference in response derives from the dependence of superflow on the gradient of the macroscopic wavefunction phase, a fact that also leads to quantised circulation and the existence of vortices and persistent currents \cite{leggett2006quantum}. Experimentally, the absence of transverse response is striking with a well-known manifestation being the Hess-Fairbank effect (analogous to the Meissner effect in superconductors). Following this logic, the standard definition of the superfluid fraction is given by finding what part of the system responds to longitudinal but not transverse perturbations\cite{pitaevskii2003bose, baym1968microscopic, nozieres1999theory, griffin1993excitations, keeling2011superfluid, carusotto2013quantum, leggett1998superfluid, rousseau2014superfluid}.
Such a definition of superfluidity is equivalent to the use of the Meissner effect to distinguish a superconductor from a material with low resistance.

\begin{figure}[b]
\centering
\inc[width=\columnwidth]{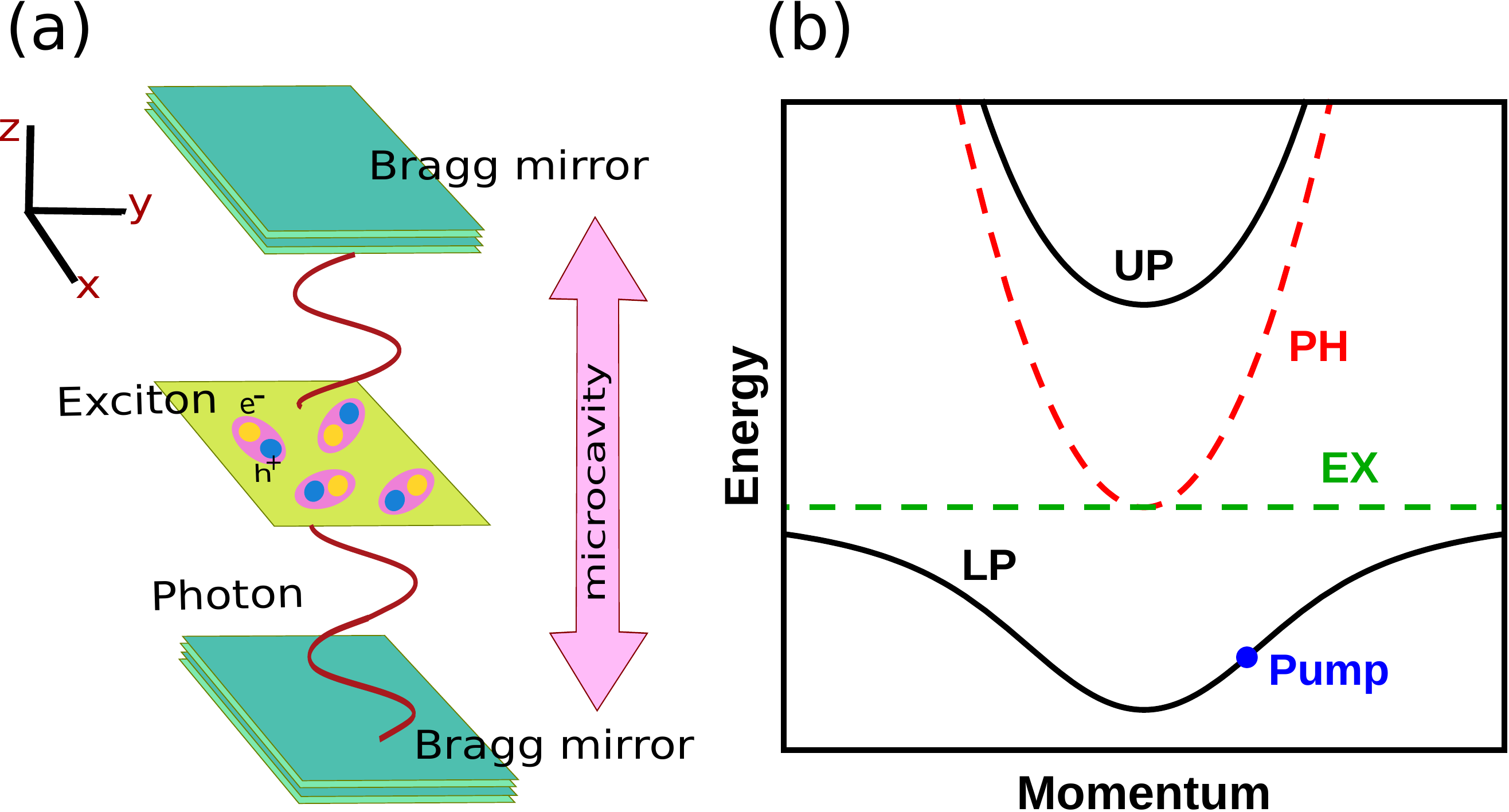}
\caption{\textbf{Polaritons in semiconductor microcavities.}  \textbf{(a)}
  Polaritons are quasiparticles formed when cavity photons, which are massive due 
  to confinement in the $z$ direction between two Bragg mirrors, interact strongly with excitons confined in a quantum
  well. Polaritons are free to move in the two-dimensional plane
  perpendicular to their confinement. \textbf{(b)} The excitonic
  dispersion (dashed green) is approximately constant compared to the
  photonic (dashed red) due to the much larger exciton mass. Strong
  coupling leads to anticrossing and the formation of upper and lower
  polariton branches (solid black). Polaritons interact because of 
  their excitonic component, while their photonic part causes decay
  and the need for an external drive. A coherent laser pump resonantly tuned to the
  polariton dispersion is marked by a blue dot.}
\label{fig:polaritons}
\end{figure}

Driven-dissipative systems present new challenges in the study of
superfluidity as they do not usually thermalise and it is unclear whether the 
effects seen in equilibrium will all continue to apply\cite{keeling2011superfluid,
keeling2009condensed,cancellieri2010superflow,carusotto2013quantum,janot2013superfluid,
gladilin2016normal}. Examples of such systems are numerous, including Bose-Einstein condensates of photons 
\cite{klaers2010bose, marelic2015experimental},
cold atoms coupled to photonic modes in optical cavities,
\cite{ritsch2013cold} and cavity arrays
\cite{hartmann2008quantum, carusotto2009fermionized,noh2016quantum}. Much recent research in this area has been focused on
microcavity-polaritons \cite{carusotto2013quantum, deng2010exciton, keeling2011exciton}, which are bosonic
quasiparticles made of quantum well excitons strongly coupled to
cavity photons (see Fig.~\ref{fig:polaritons}). Polariton experiments
have observed a number of effects usually associated with superfluidity, such as the
suppression of scattering for flow past a defect \cite{amo2009superfluidity,
  amo2009collective, berceanu2015multicomponent}, quantised vortices
\cite{lagoudakis2008quantized}, and metastable persistent currents
\cite{sanvitto2010persistent}.

The way a polaritonic system is pumped affects its excitation spectrum
and so is likely to alter its superfluid properties. While
experimentally uninvestigated, the transverse response of
incoherently pumped polaritons has been calculated using a
Keldysh path integral method \cite{keeling2011superfluid,keeling2017superfluidity}, and it was
found that a finite superfluid fraction can exist despite 
the system being out-of-equilibrium.
Fundamentally, this was a direct consequence
of the gaplessness of the diffusive excitation spectrum
\cite{szymanska2006nonequilibrium}.  This conclusion seems to hold even when going beyond a linearised theory, and considering the full nonlinear dynamics which lead to a Kardar-Parisi-Zhang (KPZ) equation predicting the absence of algebraic order~\cite{altman2015two}.
Superfluidity is
endangered by repulsive forces between vortices in the KPZ-ordered phase at 
large distances \cite{sieberer2016lattice, wachtel2016electrodynamic} --- however this can only be seen at scales too
large to be relevant in current experiments. 

On the other hand,
coherently driven polaritons below the threshold of optical parametric 
oscillation (OPO), which we consider in this work, are quite different. 
Here, the system inherits its macroscopic phase from
the pump, resulting in a gapped excitation spectrum\cite{ciuti2005quantum}. 
This phase fixing also inhibits superfluid effects such as the formation of 
vortices and solitons. However, it is notable that nearly dissipationless 
flow has been observed in this system and described as evidence of superfluidity \cite{amo2009superfluidity}.

In this Article, we calculate the
longitudinal and transverse response functions for coherently pumped
polaritons below the OPO threshold. We consider the case of a continuous and homogeneous pump -- a regime in which dissipationless flow was reported in experiment\cite{amo2009superfluidity}. We find these response functions to be equal, meaning that no part of the
system responds to perturbations like a superfluid. Furthermore, we
discover that a fraction of the system cannot be classified as `normal' or
`superfluid' as it does not respond to either longitudinal or
transverse forces.
We show this rigidity to follow from external phase fixing. Because the excitation spectrum is gapped, the KPZ nonlinearity is not relevant and our result holds in the thermodynamic limit~\cite{altman2015two,keeling2017superfluidity}.\\

\noindent\textbf{\large Results \normalsize}\\ 
\noindent\textbf{Gapped excitation spectrum.} We study
microcavity-polaritons, which are two-dimensional bosonic
quasiparticles that result from the strong coupling of quantum well
excitons to cavity photons (see Fig.~\ref{fig:polaritons}).  They have
a very small mass ($\sim 10^{-4} m_{\mathrm{e}}$, where $m_{\mathrm{e}}$ is the mass of an
electron) which allows the formation of macroscopically ordered states
at high temperatures ($\sim 20$K in GaAs\cite{kasprzak2006bose}, and up
to room temperature in organic materials \cite{daskalakis2014nonlinear}).

More specifically, we are interested in coherently pumped microcavities,
where laser excitation resonant or near-resonant with the lower 
polariton dispersion maintains a steady state against polariton decay,
leading to the formation of a macroscopically occupied state at the
pump frequency and momentum ($\omega_{\mathrm{p}}$ and $\bm{\mathrm{k}}_{\mathrm{p}}=(k_{\mathrm{p}},0)$).
\cite{stevenson2000continuous} 
In such systems there is no spontaneous symmetry breaking as the phase is fixed by the pump.

As we are interested in low energies, we can ignore the upper polariton 
dispersion and write the Hamiltonian of the system in terms of lower polariton 
operators, $\hat{a}$, coupled to bosonic decay bath modes, $\hat{A}$ \cite{dunnett2016keldysh}:
\beqy
\label{Hamiltonian}
\hat{H}&=&\sum_{\bm{\mathrm{k}}}\epsilon_{\bm{\mathrm{k}}}\hat{a}_{\bm{\mathrm{k}}}^\dagger\hat{a}_{\bm{\mathrm{k}}}+\frac{F_{\mathrm{p}}}{\sqrt{2}}(\hat{a}_{\bm{0}}^\dagger+\hat{a}_{\bm{0}})\\[0.5em]
&&+\frac{V}{2}\sum_{\bm{\mathrm{k}},\bm{\mathrm{k}}',\bm{\mathrm{q}}}\hat{a}_{\bm{\mathrm{k}}-\bm{\mathrm{q}}}^\dagger\hat{a}_{\bm{\mathrm{k}}'+\bm{\mathrm{q}}}^\dagger\hat{a}_{\bm{\mathrm{k}}}\hat{a}_{\bm{\mathrm{k}}'}+\sum_{\bm{\mathrm{p}}}\omega^{\mathrm{A}}_{\bm{\mathrm{p}}}\hat{A}_{\bm{\mathrm{p}}}^\dagger\hat{A}_{\bm{\mathrm{p}}}\nonumber\\[0.5em]
&&+\sum_{\bm{\mathrm{k}},\bm{\mathrm{p}}}\zeta_{\bm{\mathrm{k}},\bm{\mathrm{p}}}(\hat{a}_{\bm{\mathrm{k}}}^\dagger\hat{A}_{\bm{\mathrm{p}}}+\hat{A}_{\bm{\mathrm{p}}}^\dagger\hat{a}_{\bm{\mathrm{k}}}),\nonumber
\eeqy
where momentum arguments are with respect to the pump
frame.  As a result, the lower polariton dispersion takes the form
$\epsilon_{\bm{\mathrm{k}}}=(\bm{\mathrm{k}}+\bm{\mathrm{k}}_{\mathrm{p}})^2/2m^*$, where we have used
a quadratic approximation.  Other quantities appearing in this expression
are: $F_{\mathrm{p}}$, the amplitude of the pump, $V$, the
polariton-polariton interaction strength, $\omega^{\mathrm{A}}_{\bm{\mathrm{p}}}$
the dispersion of the bath modes, and $\zeta_{\bm{\mathrm{k}},\bm{\mathrm{p}}}$,  the
coupling between the polaritons and the bath modes. While  for some range of detunings and pump intensities this system becomes bistable,  here we consider the monostable state at low pump intensity.

\begin{figure}[b]
\centering
\inc[width=\columnwidth]{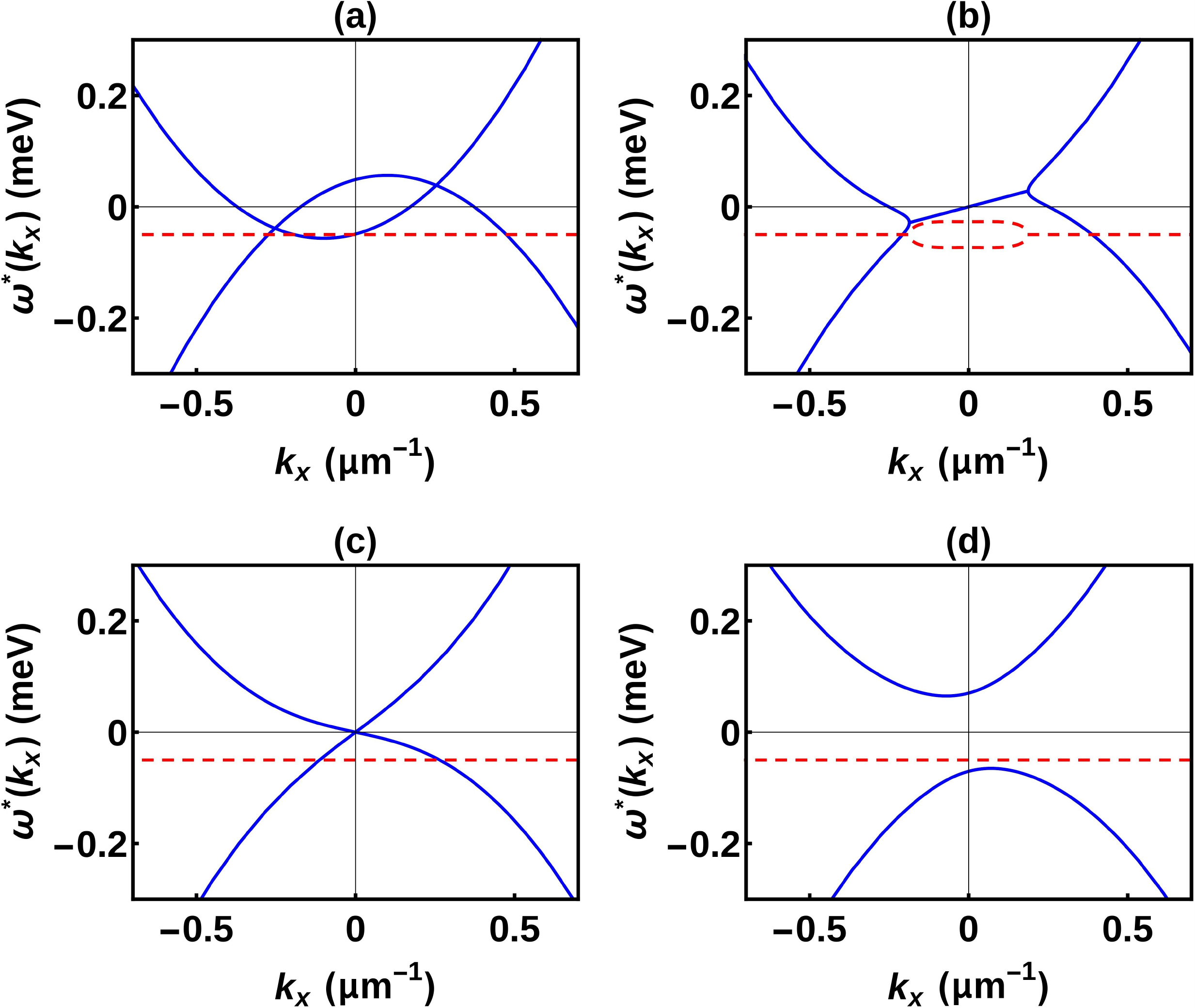}
\caption{\textbf{The excitation spectrum at different
      polariton densities.} The real (solid blue) and the imaginary
    (dashed red) parts of the spectrum in the frame of the pump
    momentum, which is $k_{\mathrm{p}}=0.1\mathrm{\upmu m^{-1}}$ in the
    $x$-direction, for $k_y=0$.  The pump is blue detuned by
    $0.05\mathrm{meV}$ with respect to the bare lower polariton
    dispersion. In \textbf{(a)} and \textbf{(b)} the densities are
  $\abs{\psi_0}^2=0.2\mathrm{\upmu m}^{-2}$ and
  $\abs{\psi_0}^2=9.3\mathrm{\upmu m}^{-2}$ respectively. In
  \textbf{(c)} the density is $\abs{\psi_0}^2=20.0\mathrm{\upmu m}^{-2}$
  and the pump comes into resonance with the interaction shifted lower
  polariton dispersion,
  $\Delta_{\mathrm{p}}=\omega_{\mathrm{p}}-\epsilon_{\bm{0}}-V\abs{\psi_0}^2=0$, at which point the real part takes the linear form of the Bogoliubov spectrum. Here, and in
  \textbf{(d)} where $\abs{\psi_0}^2=30.9\mathrm{\upmu m}^{-2}$ the
  Landau criterion is fulfilled in the real part. However, it is
  significant that the imaginary part is always gapped.}
\label{fig:spectra}
\end{figure}

The excitation spectrum of this system has been studied previously
\cite{ciuti2005quantum,dunnett2016keldysh,carusotto2004probing}, and
is given by
\beq
\label{spectrum}
\omega^{*,\pm}_{\bm{\mathrm{k}}}=\frac{\alpha^+_{\bm{\mathrm{k}}}-\alpha^-_{\bm{\mathrm{k}}}}{2}-\mathrm{i}\kappa\pm\frac{1}{2}\sqrt{(\alpha^+_{\bm{\mathrm{k}}}+\alpha^-_{\bm{\mathrm{k}}})^2-4V^2\abs{\psi_0}^4}, 
\eeq
where $\psi_0$ is the mean-field solution,
$\alpha^{\pm}_{\bm{\mathrm{k}}}=\epsilon_{\pm\bm{\mathrm{k}}}-\omega_{\mathrm{p}}+2V\abs{\psi_0}^2$ and
$\kappa$ is a decay constant derived from integrating out the bath.
Here, we wish to emphasise a point somewhat neglected in previous
work, which has concentrated on the gaplessness of the real part of
the excitation spectrum for specific blue detuning
($\Delta_{\mathrm{p}}=\omega_{\mathrm{p}}-\epsilon_{\bm{0}}-V\abs{\psi_0}^2=0$), at which it takes a linear Bogoliubov-like form near $\omega, k=0$. It has been
noted that in this regime the real part appears to fulfil the 
Landau criterion for superfluidity, and this fact has been used to explain the 
observation of dissipationless flow \cite{amo2009superfluidity}. However, we 
note that unlike in equilibrium systems, for which the criterion was derived, 
the excitation spectrum here is complex and, due to phase fixing by the pump, 
is gapped, except at exactly the pump strength where a parametric instability first occurs. In general this gap is found in the imaginary part (see Fig.~\ref{fig:spectra}).
In fact, it has been shown that scattering in these systems can only be reduced, not completely eliminated \cite{cancellieri2010superflow}. Furthermore, 
while approximately dissipationless flow can be explained by the real part of 
the excitation spectrum, a gapped spectrum may have important consequences 
for superfluidity more generally. Indeed, the limits of using the Landau 
criterion alone to interpret superfluidity in driven-dissipative systems 
can be seen from the incoherent case, where the diffusive excitation 
spectrum\cite{szymanska2006nonequilibrium} does not fulfil it at all. 
In this context, a new generalised criterion in terms of the complex wave vector was formulated to explain dissipationless flow\cite{wouters2010superfluidityPRL}.\\\\

\begin{figure}[t]
\centering
\inc[width=\columnwidth]{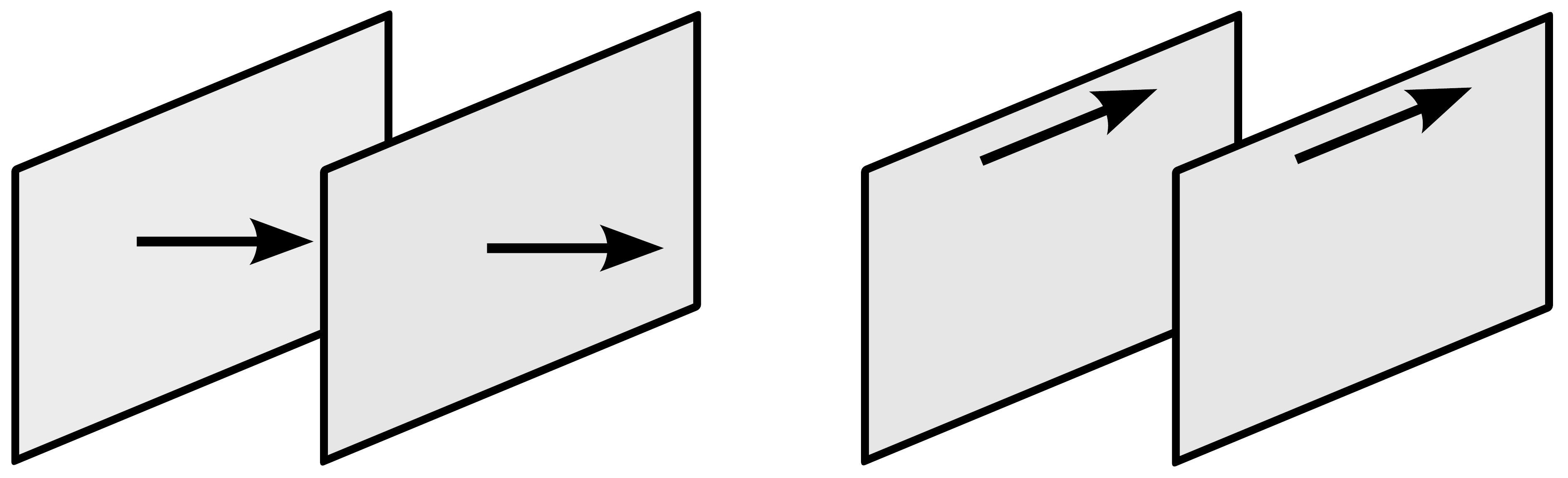}
\caption{\textbf{Longitudinal vs. transverse response.} A current flowing in response to a perturbing force may originate through longitudinal or transverse actions, corresponding to pushing and shearing respectively. A superfluid responds to longitudinal but not transverse perturbations\cite{pitaevskii2003bose, baym1968microscopic, nozieres1999theory, griffin1993excitations, keeling2011superfluid, carusotto2013quantum, leggett1998superfluid, rousseau2014superfluid}.}
\label{fig:longandtran}
\end{figure}

\noindent\textbf{Sum-rules and anisotropy.} To study whether the phase
fixing and gapped spectrum in coherently pumped polaritons affects
their superfluid properties, we calculate the static current-current
response function. More specifically, for a perturbation described by the Hamiltonian $\hat{H}=\sum_{\bm{\mathrm{q}}}\hat{\bm{\mathrm{j}}}(\bm{\mathrm{q}}).\bm{\mathrm{f}}(\bm{\mathrm{q}})$, where $\bm{\mathrm{f}}$ is the perturbating force, this response function describes the tendency for a particle current to flow as a result of that
force, $j_i(\bm{\mathrm{q}})=\chi_{ij}(\bm{\mathrm{q}})f_j(\bm{\mathrm{q}})$,  where $i,j$ refer to directions in the $xy$ plane. This function is given by the correlator of two current operators
\cite{altland2010condensed}, $\chi_{ij}(\bm{\mathrm{q}})=-\mathrm{i}\langle
\hat{j}_i(\bm{\mathrm{q}})\hat{j}_j(-\bm{\mathrm{q}})\rangle$, and can be separated into
longitudinal and transverse parts (i.e. the response to pushing or shearing the polaritons as shown in Fig.~\ref{fig:longandtran}). A perfect superfluid has only the former. Mathematically, this decomposition can be made by
considering a diagonal component of the response tensor, $\chi_{ii}$, and varying the order in which the components of the momentum vector are
taken to zero\cite{pitaevskii2003bose,huang1995bose}. Taking
the transverse momentum (i.e. the component perpendicular to $i$)  to zero before the longitudinal  momentum (component parallel to $i$) gives
the longitudinal response function, and vice versa.

In systems where particle number is conserved, the f-sum rule holds.
\cite{pitaevskii2003bose} This identity follows directly from continuity of current,
\beq
[\hat{H},\hat{\rho}(\bm{\mathrm{q}})]=-\bm{\mathrm{q}}.\hat{\bm{\mathrm{j}}}(\bm{\mathrm{q}}),
\label{continuity}
\eeq
and allows the longitudinal response
function in the long wavelength limit to be identified
with the total density, $\rho=m\chi^{\mathrm{L}}(\bm{\mathrm{q}}\rightarrow\bm{0})$. In addition, using linear response theory, one
can identify the normal density as the transverse response function in
the same limit \cite{pitaevskii2003bose,griffin1993excitations}, $\rho_{\mathrm{n}}=m\chi^{\mathrm{T}}(\bm{\mathrm{q}}\rightarrow\bm{\mathrm{0}})$, and calculate the
superfluid density as the difference between the two limiting responses
\cite{pitaevskii2003bose, baym1968microscopic, nozieres1999theory,
  griffin1993excitations, keeling2011superfluid}, $\rho_{\mathrm{s}}=m(\chi^{\mathrm{L}}(\bm{\mathrm{q}}\rightarrow\bm{0})-\chi^{\mathrm{T}}(\bm{\mathrm{q}}\rightarrow\bm{0}))$. In
driven-dissipative systems however, particle conservation does not
hold and one should not expect the f-sum rule or, by extension the above expression for $\rho_{\mathrm{s}}$, to hold either. While
previous work on incoherently pumped polaritons nevertheless found
that the f-sum rule did hold \cite{keeling2011superfluid}, coherently
pumped systems face further complications. In particular, for the
total density of the system to equal the sum of the normal and
superfluid components, $\rho=\rho_{\mathrm{n}}+\rho_{\mathrm{s}}$, the system must be Galilean invariant
\cite{leggett2006quantum}. It is clear that Galilean invariance is broken for a coherently pumped
system, as polaritons are injected at a fixed momentum, picking out a
special frame of reference through the pumping term in the
Hamiltonian (Eq. (\ref{Hamiltonian})). In addition,
  coherently pumped polaritons are anisotropic, meaning that the
  longitudinal and transverse response functions are tensors, not
  scalars. Given these subtleties, a relation between the response
  functions and densities is unknown.  Thus, instead we determine
  whether any part of the system responds to perturbations in a manner
  characteristic of a superfluid, as defined by the superfluid
  response:
\beq
\label{superfluid}
\lim_{\bm{\mathrm{q}}\rightarrow\bm{0}}\left(\chi^{\mathrm{S}}_{ij}(\bm{\mathrm{q}})\right)=\lim_{\bm{\mathrm{q}}\rightarrow\bm{0}}\left(\chi^{\mathrm{L}}_{ij}(\bm{\mathrm{q}})-\chi^{\mathrm{T}}_{ij}(\bm{\mathrm{q}})\right),
\eeq
where $\chi^{\mathrm{L}}_{ij}$ and $\chi^{\mathrm{T}}_{ij}$ are the (anisotropic)
longitudinal and transverse response functions. We will not however associate these with densities. \\

\noindent\textbf{Response function.} In order to calculate the static current-current response function we utilise a path integral method where the perturbation is modelled by source fields\cite{keeling2011superfluid}. Here we present the results of this calculation, with details of the derivation using Keldysh field theory given in the Methods.

Each term in the response function contains two
factors of the momentum vertex,
\beq
\bm{\upgamma}(\bm{\mathrm{k}}+\bm{\mathrm{q}},\bm{\mathrm{k}})=\frac{1}{2m^*}
\bmat
2k_{\mathrm{p}}+2k_x+q_x\\[0.5em]
2k_y+q_y
\emat,
\eeq
which describes the bare coupling between the source fields and excitations
carrying momentum $\bm{\mathrm{q}}$. Because the pump breaks rotational invariance the vertex is anisotropic. Without loss of generality
we consider the pump wavevector to be in the $x$ direction.

The full response can be written in terms of two components, one due to the mean-field and the other due to fluctuations:
\beq
\label{chi}
\chi_{ij}(\bm{\mathrm{q}})=\chi_{ij}^{\mathrm{mf}}(\bm{\mathrm{q}})+\chi_{ij}^{\mathrm{fl}}(\bm{\mathrm{q}}).
\eeq
The mean-field term is,
\begin{equation}
\label{coeff1}
\chi_{ij}^{\mathrm{mf}}(\bm{\mathrm{q}})=\sum_{\sigma,\sigma'\in\pm}c^{\mathrm{mf}}_{\sigma,\sigma'}(\bm{\mathrm{q}})\gamma_i(\sigma\bm{\mathrm{q}})\gamma_j(\sigma'\bm{\mathrm{q}}),
\end{equation}
where the coefficient $c^{\mathrm{mf}}_{\sigma,\sigma'}(\bm{\mathrm{q}})$ is given in Supplementary Note 1, and the fluctuations term is:
\beqy
\chi_{ij}^{\mathrm{fl}}(\bm{\mathrm{q}})&=&\frac{\mathrm{i}}{4}\int\frac{d^2\bm{\mathrm{k}}}{(2\uppi)^2}\bigg\{\mathrm{i}\Tr[\mathcal{M}_{ij}^{\bm{\mathrm{q}}}]+\int\frac{d\omega}{2\uppi}\bigg(\Tr[\mathcal{D}^{\mathrm{K}}_{\omega,\bm{\mathrm{k}}}\mathcal{M}_{ij}^{\bm{\mathrm{q}}}]\nonumber\\[0.5em]
&&-\Tr\Big[\mathcal{D}^{\mathrm{R}}_{\omega,\bm{\mathrm{k}}+\bm{\mathrm{q}}}\mathcal{A}_i^{\bm{\mathrm{k}}+\bm{\mathrm{q}},\bm{\mathrm{k}}}\mathcal{D}^{\mathrm{K}}_{\omega,\bm{\mathrm{k}}}\mathcal{B}_j^{\bm{\mathrm{k}},\bm{\mathrm{k}}+\bm{\mathrm{q}}}\\[0.5em]
&&\quad\;{}+\mathcal{D}^{\mathrm{K}}_{\omega,\bm{\mathrm{k}}+\bm{\mathrm{q}}}\mathcal{A}_i^{\bm{\mathrm{k}}+\bm{\mathrm{q}},\bm{\mathrm{k}}}\mathcal{D}^{\mathrm{A}}_{\omega,\bm{\mathrm{k}}}\mathcal{B}_j^{\bm{\mathrm{k}},\bm{\mathrm{k}}+\bm{\mathrm{q}}}\Big]\bigg)\bigg\}\nonumber
\eeqy
where the retarded, advanced, and Keldysh Green's functions are defined by $\mathcal{D}^{\mathrm{R}}=-\mathrm{i}\theta(t-t')\langle[\Psi(t),\Psi^{\dagger}(t')]\rangle$, $\mathcal{D}^{\mathrm{A}}=\mathrm{i}\theta(t'-t)\langle[\Psi(t),\Psi^{\dagger}(t')]\rangle$, and $\mathcal{D}^{\mathrm{K}}=-\mathrm{i}\langle\{\Psi(t),\Psi^{\dagger}(t')\}\rangle$. The Green's functions, as well as
$\mathcal{A}_i^{\bm{\mathrm{k}}+\bm{\mathrm{q}},\bm{\mathrm{k}}}$,
$\mathcal{B}_i^{\bm{\mathrm{k}},\bm{\mathrm{k}}+\bm{\mathrm{q}}}$, and
$\mathcal{M}_{ij}^{\bm{\mathrm{q}}}$, are $2\times2$ matrices in Nambu space (i.e., the space of particle creation and anihilation operators). The 
matrices consist of a series of
different combinations of momentum vertices:
\begin{flalign}
\label{coeff2}
\mathcal{A}_i^{\bm{\mathrm{k}}+\bm{\mathrm{q}},\bm{\mathrm{k}}}/\mathcal{B}_i^{\bm{\mathrm{k}}+\bm{\mathrm{q}},\bm{\mathrm{k}}}&=\frac{1}{2}(\mathbb{1}+\hat{\sigma}_z)\gamma_i(\bm{\mathrm{k}}+\bm{\mathrm{q}},\bm{\mathrm{k}})&\\
&\quad+\frac{1}{2}(\mathbb{1}-\hat{\sigma}_z)\gamma_i(-\bm{\mathrm{k}}-\bm{\mathrm{q}},-\bm{\mathrm{k}})&\nonumber\\
&\quad+\sum_{\sigma\in\pm}\mathcal{C}^{\mathcal{A}/\mathcal{B}}_{\sigma}(\bm{\mathrm{q}})\gamma_i(\sigma\bm{\mathrm{q}}),&\nonumber
\end{flalign}
\begin{flalign}
\label{coeff3}
\mathcal{M}_{ij}^{\bm{\mathrm{q}}}&=\sum_{\sigma,\sigma'\in\pm}\mathcal{C}_{\sigma,\sigma'}^{\mathcal{M}}(\bm{\mathrm{q}})\gamma_i(\sigma\bm{\mathrm{q}})\gamma_j(\sigma'\bm{\mathrm{q}})&
\end{flalign}
where $\mathcal{C}^{\mathcal{A}/\mathcal{B}}_{\sigma}(\bm{\mathrm{q}})$ and $\mathcal{C}^{\mathcal{M}}_{\sigma,\sigma'}(\bm{\mathrm{q}})$ are $2\times2$ matrices in Nambu space, and $\hat{\sigma}_z$ is the third Pauli matrix. These coefficients are given in Supplementary Note 1. 

As discussed further below, it is important to observe that each of these terms involves $1/\det[(D^{\mathrm{R}})^{-1}(\omega=0,\bm{\mathrm{q}})]$.  One may also note that the total response function can be divided into two sets of contributions.
Those involving the response of the condensate appearing through the coupling $\bm{\upgamma}(\bm{0},\bm{\mathrm{q}})=(2\bm{\mathrm{k}}_{\mathrm{p}}+\bm{\mathrm{q}})/2m^*$, and those involving coupling to excitations through $\bm{\upgamma}(\bm{\mathrm{k}}+\bm{\mathrm{q}},\bm{\mathrm{k}})=(2\bm{\mathrm{k}}_{\mathrm{p}}+2\bm{\mathrm{k}}+\bm{\mathrm{q}})/2m^*$.\\\\
\textbf{Superfluid response.} To quantify the superfluid behaviour of the system we need to take the long wavelength limit of the response function. In order for the longitudinal and transverse responses to differ --- that is, for the order in which $q_x$ and $q_y$ are taken to zero to matter in any way ---  there needs to be singular behaviour as $\bm{\mathrm{q}}\rightarrow\bm{0}$. Each term in Eq. (\ref{chi}) takes the ultimate form
\beq
\frac{h(\bm{\mathrm{q}})}{\det[(D^{\mathrm{R}})^{-1}(\omega=0,\bm{\mathrm{q}})]},
\eeq
where $h(\bm{\mathrm{q}})$ is some polynomial in $\bm{\mathrm{q}}$.  Singular behaviour requires the denominator of this expression, the static inverse retarded Green's function, to vanish as $\bm{\mathrm{q}} \to \bm{\mathrm{0}}$.  One may see that the requirement for this to happen is directly related to the gaplessness of the excitation spectrum.  The spectrum, $\omega^*(\bm{\mathrm{k}})$, in Eq. (\ref{spectrum}) is defined by
\beq
\label{spectrum2}
\det[(D^{\mathrm{R}})^{-1}(\omega^*(\bm{\mathrm{k}}),\bm{\mathrm{k}})]=0.
\eeq
If the spectrum is gapless,  i.e. if $\omega^*(\bm{\mathrm{k}}\rightarrow\bm{0})=0$, then the static inverse retarded Green's function will vanish as $\bm{\mathrm{q}}$ goes to zero,
and there can be a singular dependence of the response function on momentum.
As the excitation spectrum of coherently pumped polaritons is always gapped this instead gives a finite value for the static inverse retarded Green's function, so there cannot be any singular terms. Consequently, the longitudinal and transverse response functions are equal and Eq. (\ref{superfluid}) shows that no part of the system responds to perturbations like a superfluid.\\
\begin{figure}[b]
\centering
\inc[width=\columnwidth]{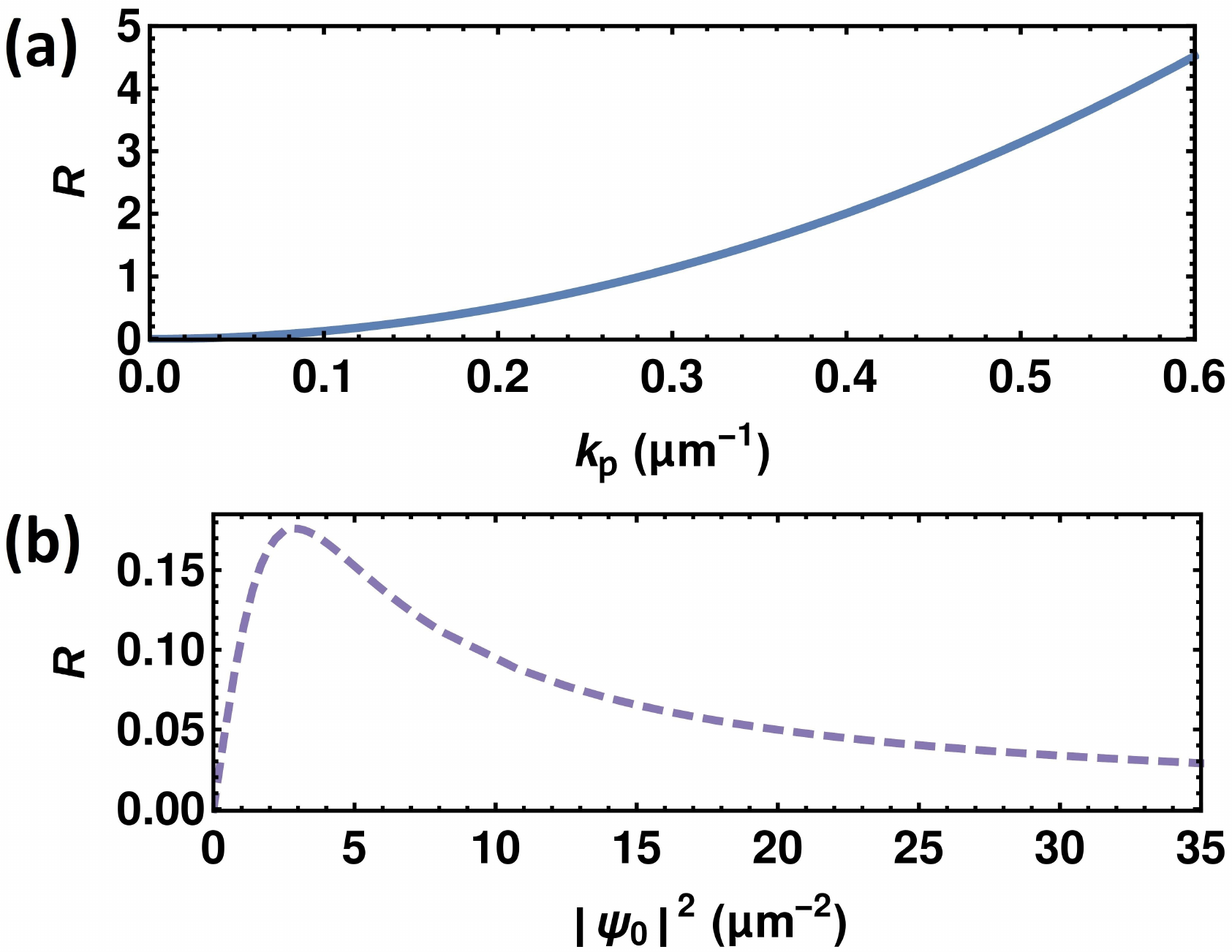}
\caption{\textbf{Response at zero detuning.}
    The long wavelength response, normalised to the mean-field
    density, $R=m^*\chi_{xx}(\bm{\mathrm{q}}\rightarrow\bm{0})/\abs{\psi_0}^2$, as a function \textbf{(a)} of pump momentum $k_{\mathrm{p}}$  at
    $\abs{\psi_0}^2=6.9\mathrm{\upmu m}^{-2}$, and \textbf{(b)} the
    mean-field density $\abs{\psi_0}^2$ at
    $k_{\mathrm{p}}=0.1\mathrm{\upmu m}^{-1}$. The detuning is zero
    in both cases.}
\label{fig:ratios}
\end{figure}
\begin{figure}[t]
\centering
\inc[width=\columnwidth]{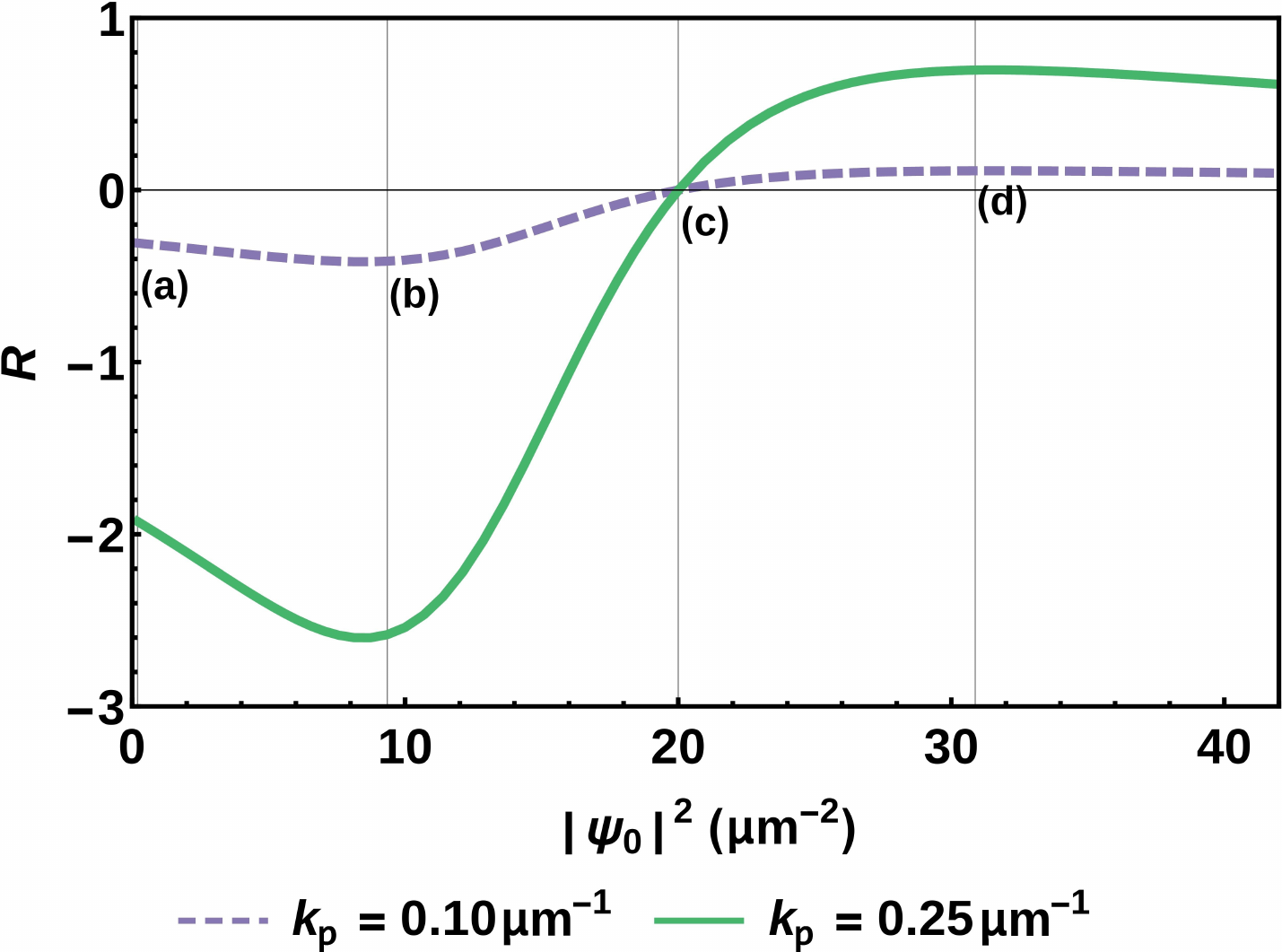}
\caption{\textbf{Response at finite detuning.}
    Total long wavelength response for $k_{\mathrm{p}}=0.1\mathrm{\upmu m^{-1}}$ (dashed
    purple) and $k_{\mathrm{p}}=0.25\mathrm{\upmu m^{-1}}$ (solid green) normalised with respect
    to the mean-field density, $R=m^*\chi_{xx}(\bm{\mathrm{q}}\rightarrow\bm{0})/\abs{\psi_0}^2$. Vertical lines marked (a-d) correspond to the
  excitation spectra in Fig.~\ref{fig:spectra}. 
  For (a) and (b) the response is negative, and goes to zero
  close to (c) where the excitation spectrum is of the linear Bogoliubov form (the mean-field component, Eq. (\ref{chi limit}),
  is exactly zero at (c) on account of an imaginary mean-field, but
  the fluctuations are small and finite), before becoming positive,
  peaking around (d), and tailing off similarly to the resonant case
  in Fig.~\ref{fig:ratios}. Larger $k_{\mathrm{p}}$ leads to a larger response
  to perturbations. Note that there is no change in response associated with the speed of sound, 
  given by $c_{\mathrm{s}}=\sqrt{V\abs{\psi_0}^2/m^*}$, which corresponds to the momentum $k=0.1\mathrm{\upmu m^{-1}}$ 
  when $\abs{\psi_0}^2=6.1\mathrm{\upmu m}^{-2}$ and $k=0.25\mathrm{\upmu m^{-1}}$ when $\abs{\psi_0}^2=38.1\mathrm{\upmu m}^{-2}$.}
\label{fig:response}
\end{figure}

\noindent\textbf{Rigid state.} In fact our result is more profound. In
the case where $\bm{\mathrm{k}}_{\mathrm{p}}=\bm{0}$, the response function is exactly zero
in the long wavelength limit: that is, the system does not
respond to either longitudinal or transverse perturbations. This
implies that the coherently pumped system forms a rigid state more
akin to a solid than a fluid in that it has a finite density but shows zero response to
small perturbations. Indeed, while it is already known that phase
fixing prevents the formation of vortices~\cite{pigeon2011hydrodynamic}, we can see that its effect is even more severe in its restriction of the response of the system.
(If the pump is restricted in time or in space, then there can be free phase evolution in those times or places without a pump, allowing vortex formation to be recovered~\cite{amo2011polariton,nardin2011hydrodynamic}.)

When the pump is at non-zero wavevector, $\bm{\mathrm{k}}_{\mathrm{p}}\neq\bm{0}$, the behaviour is different.  While the state remains rigid to perturbations (in the sense that its momentum continues to be locked to the pump) the net current can change, by modifying the occupation of the state at $\bm{\mathrm{k}}_{\mathrm{p}}$:  i.e. if the coherent state has a finite momentum, a change in the occupation of that state will change the total current, allowing a non-zero response to the applied force.  Mathematically, this occurs because
at finite $\bm{\mathrm{k}}_{\mathrm{p}}$ the perturbation
couples to the pump state through the momentum vertex
$\bm{\upgamma}(\bm{0},\bm{0})=\bm{\mathrm{k}}_{\mathrm{p}}/2m^*$, and so a force can change the amplitude of the state at $\bm{\mathrm{k}}_{\mathrm{p}}$. This change manifests as a
finite normal response with quadratic dependence on
$k_{\mathrm{p}}$. (By `normal' we mean that the longitudinal and transverse responses are equal.) As a result, the mean-field part of the response, which is orders of magnitude larger than the part due to fluctuations, takes the form
\beq
\label{chi limit}
\chi_{ij}^{(0)}(\bm{\mathrm{q}}\rightarrow\bm{0})=-\delta_{xi}\delta_{xj}\frac{k_{\mathrm{p}}^2\abs{\psi_0}^2(\bar{\psi}_0+\psi_0)}{m^{*2}(F_{\mathrm{p}}-V\abs{\psi_0}^2(\bar{\psi}_0+\psi_0))}.   
\eeq
This expression reports how a perturbation changes the occupation of the pump state, which, for finite pump momentum, changes the overall current.
To see this in detail, one can start with the time-independent Gross-Pitaevskii
  equation (GPE) for a bosonic field $\psi$ coupled to a current
  through the perturbation $\bm{\mathrm{f}}(\bm{\mathrm{x}})$:
\begin{multline}
  \bigg(-\frac{\nabla^2}{2m}-\omega_{\mathrm{p}}-\mathrm{i}\kappa+V\abs{\psi(\bm{\mathrm{x}})}^2\bigg)\psi(\bm{\mathrm{x}})=-F_{\mathrm{p}}\e^{\mathrm{i}\bm{\mathrm{k}}_{\mathrm{p}}.\bm{\mathrm{x}}}\\
  -\frac{\mathrm{i}}{2m}\left[\bm{\mathrm{f}}(\bm{\mathrm{x}}).\nabla\psi(\bm{\mathrm{x}})+\nabla.(\bm{\mathrm{f}}(\bm{\mathrm{x}})\psi(\bm{\mathrm{x}}))\right],
\end{multline}
and find how the occupation of the pump state changes due to this perturbation. Because we are interested in the $\bm{\mathrm{q}}\rightarrow\bm{0}$ limit, we take $\bm{\mathrm{f}}$ to be independent of position, so this equation reduces to
\begin{multline}
\bigg(-\frac{\nabla^2}{2m}-\omega_{\mathrm{p}}-\mathrm{i}\kappa+
  V\abs{\psi(\bm{\mathrm{x}})}^2\bigg)\psi(\bm{\mathrm{x}})\\
=-F_{\mathrm{p}}\e^{\mathrm{i}\bm{\mathrm{k}}_{\mathrm{p}}.\bm{\mathrm{x}}}-\bm{\mathrm{f}}.
 \frac{\mathrm{i}\nabla}{m}\psi(\bm{\mathrm{x}}).
\end{multline}
Rewriting
  $\psi(\bm{\mathrm{x}})=(\psi_0+\phi(\bm{\mathrm{x}}))\e^{\mathrm{i}\bm{\mathrm{k}}_{\mathrm{p}}.\bm{\mathrm{x}}}$, where
  $\psi_0\e^{\mathrm{i}\bm{\mathrm{k}}_{\mathrm{p}}.\bm{\mathrm{x}}}$ is the mean-field satisfying the
  original GPE (i.e. with $\bm{\mathrm{f}}=\bm{0}$), and removing any terms higher than first order in
  $\phi(\bm{\mathrm{x}})$ and $\bm{\mathrm{f}}$ leads to the equation
\begin{multline}
  -\frac{\nabla^2}{2m}\phi(\bm{\mathrm{x}})-\mathrm{i}\bm{\mathrm{k}}_{\mathrm{p}}.\frac{\nabla}{m}\phi(\bm{\mathrm{x}})+V\psi_0^2\phi^*(\bm{\mathrm{x}})\\
  +\left(\frac{k_{\mathrm{p}}^2}{2m}-\omega_{\mathrm{p}}-\mathrm{i}\kappa+2V\abs{\psi_0}^2\right)\phi(\bm{\mathrm{x}})=\frac{\bm{\mathrm{f}}.\bm{\mathrm{k}}_{\mathrm{p}}}{m}\psi_0.
\end{multline}
One may check (e.g. by Fourier transforming) that the solution of this is a homogenous (position independent) function $\phi(\bm{\mathrm{x}})=\phi$. The current associated with this change of occupation is given by $\delta \bm{\mathrm{j}} = (\bm{\mathrm{k}}_{\mathrm{p}}/m)(\psi_0^\ast \phi + \phi^\ast \psi_0)$, which can be used to recover the mean-field response function through $\delta j_i=\chi^{(0)}_{ij}(\bm{\mathrm{q}}\rightarrow\bm{0})f_j$, with  $\chi^{(0)}_{ij}(\bm{\mathrm{q}}\rightarrow\bm{0})$ as given in Eq. (\ref{chi limit}).

Figure~\ref{fig:ratios} shows a quantity $R$ defined as the ratio of the long wavelength response to the mean-field density,
$R=m^*\chi_{xx}(\bm{\mathrm{q}}\rightarrow\bm{0})/\abs{\psi_0}^2$. We explore how $R$ changes with pump momentum $k_{\mathrm{p}}$ and mean-field polariton density
$\abs{\psi_0}^2$ for the parameters: $m^*=5\times10^{-5}m_\mathrm{e}$, where $m_\mathrm{e}$
 is the electron mass, $V=0.01\mathrm{meV\upmu m^2}$, and $\kappa=0.05\mathrm{meV}$. 
The ratio increases quadratically with $k_{\mathrm{p}}$,
i.e. the higher the velocity of the pump state the more 
the current changes due to the perturbation. That the 
response increases smoothly with velocity is consistent with a smooth
increase in drag force identified in a previous study
\cite{cancellieri2010superflow} of coherently pumped polaritons. There is a non monotonic dependence of $R$ on the pump intensity, with an asymptotic decrease of $R$ at large intensity.  This arises because a large pump intensity reduces the response to a weak perturbation. \\

  \noindent\textbf{Detuning.} 
  It is worth noting that blue detuning can lead to the real part of
  the excitation spectrum fulfilling the Landau criterion at the right
  density\cite{amo2009superfluidity,ciuti2005quantum,carusotto2004probing} (see
  Fig.~\ref{fig:spectra}(c)). This regime has been used to explore
  flows against defects with reduced
  dissipation\cite{amo2009superfluidity}. Fig.~\ref{fig:response}
  shows how the long wavelength response normalised to the mean-field
  density changes when we change the value of the polariton density
  for a specific blue detuning, where $m^*=5\times10^{-5}m_\mathrm{e}$, 
  $V=0.0025\mathrm{meV\upmu m^2}$, and $\kappa=0.05\mathrm{meV}$. It is notable that at the shifted
  resonance point, $\Delta_{\mathrm{p}}=\omega_{\mathrm{p}}-\epsilon_{\bm{0}}-V\abs{\psi_0}^2=0$, where the real part of the excitation spectrum takes a Bogoliubov form, the mean-field
  term in the response function, Eq. (\ref{chi limit}), becomes
  zero. The reason for this is that Eq. (\ref{chi limit}) is
  proportional to $\Re(\psi_0)$, and when $\Delta_{\mathrm{p}}=0$, $\psi_0$ is
  entirely imaginary. While the
  long wavelength response due to fluctuations is finite, it is orders
  of magnitude smaller than the mean-field contriubution at most
  densities. This suggests that, even if we take higher orders in fluctuations, there is always a pump strength for which the response goes from negative to positive and is thus strictly zero. 
  Furthermore, this pump strength is very close to the shifted resonance point at which dissipationless flow was observed in experiments. For densities above the shifted resonance point, the Landau 
  criterion is still fulfilled in the real part which is now gapped, but by contrast to the experimentally investigated regime in  Fig.~\ref{fig:spectra}(c), the response function is finite and positive, showing a gradual
  reduction with increasing density similar to the case of zero detuning
  (Fig.~\ref{fig:ratios}).

  Because there is no continuity of current (Eq. \ref{continuity}), the f-sum rule does not hold and there is no clear physical correspondence between the response functions and the density of the system. 
  As a result, the negative value of the long wavelength response function at low densities does not present a problem. Additionally, it should be noted that in a
  previous study of incoherently pumped polaritons at the mean-field
  level \cite{gladilin2016normal}, in which external potentials were
  present, it was concluded that the resultant currents in the steady
  state can change the physical picture substantially and render the
  interpretation of the superfluid and normal density fractions in terms of
  the response of the system to a vector potential unphysical. While
  the phase was free in that study and in the present work there are
  no external potentials, parallels exist with the fact that the
  coherent pumping ensures a steady state current.\\

\noindent\textbf{\large Discussion \normalsize}\\
We study, using a nonequilibrium path integral method, a system of
coherently pumped microcavity-polaritons in which the pumping is continuous, homogeneous, and below the OPO threshold, and show that the external fixing of the the macroscopic phase prevents it from
being a superfluid. This is because the
gapped spectrum affects the limiting behaviour of the current-current
response function such that the longitudinal and transverse responses
are the same. Remarkably, we also find that the system does not respond
to either longitudinal or transverse perturbations at
$\bm{\mathrm{k}}_{\mathrm{p}}=\bm{0}$, and it possesses only a normal component at finite
$\bm{\mathrm{k}}_{\mathrm{p}}$, which grows with increasing $\bm{\mathrm{\mathrm{k}}}_{\mathrm{p}}$.
Rather than a superfluid, this result suggests the existence
of a rigid state that, like a solid, has density but no current
response. The smooth growth of the normal component with
  $\bm{\mathrm{k}}_{\mathrm{p}}$ is similar to the smooth crossover in drag predicted in a previous study
  \cite{cancellieri2010superflow}. Additionally, the fraction of the system corresponding to the
  macroscopic rigid state grows faster than that of the normal component as
  pump intensity is increased.

  While blue detuning can allow the fulfilment of the Landau criterion
  in the real part, and signs of dissipationless flow have been
  observed \cite{amo2009superfluidity}, this is the only
  property associated with superfluidity that is exhibited by the rigid state, as vortices
  and persistent currents cannot form when the phase is externally
  fixed, and the superfluid response is zero. It is
  notable too that the long wavelength total current-current response function
  falls to zero very close to the experimentally investigated regime
  where the excitation spectrum takes the Bogoliubov form, which could explain
  the observed reduced scattering.

The existence of this rigid state suggests that driven-dissipative
systems allow for a richer collection of macroscopic flow properties
than equilibrium systems, and highlights the subtleties inherent in the attributes together known as superfluidity.\\ 

\noindent\textbf{\large Methods \normalsize}

\small{\noindent\textbf{Keldysh path integral.} We calculate
the  current-current response function using a Keldysh path
  integral technique \cite{keeling2011superfluid, kamenev2011field, altland2010condensed}. Starting from the
  Hamiltonian (Eq. (\ref{Hamiltonian})) and integrating out the decay
  bath, the action is given in terms of Keldysh `classical' and
  `quantum' fields, $\Psi=(\psi^{\mathrm{c}},\psi^{\mathrm{q}})$ \cite{dunnett2016keldysh}: 
\begin{flalign}
\label{S}
S[\Psi]&=\sum_{\omega,\bm{\mathrm{k}}}(\bar{\psi}^{\mathrm{c}}_{\bm{\mathrm{k}}}(\omega),\bar{\psi}^{\mathrm{q}}_{\bm{\mathrm{k}}}(\omega))
\bmat
0&D_{\bm{\mathrm{k}}}^0(\omega)\\[0.5em]
D_{\bm{\mathrm{k}}}^0(\omega)^*&2\mathrm{i}\kappa
\emat
\bmat
\psi^{\mathrm{c}}_{\bm{\mathrm{k}}}(\omega)\\[0.5em]
\psi^{\mathrm{q}}_{\bm{\mathrm{k}}}(\omega)
\emat&\\[0.5em]
&-\sum_{\omega,\omega',\nu}\sum_{\bm{\mathrm{k}},\bm{\mathrm{k}}',\bm{\mathrm{q}}}\frac{V}{2}\Bigl(\bar{\psi}^{\mathrm{c}}_{\bm{\mathrm{k}}-\bm{\mathrm{q}}}(\omega-\nu)\bar{\psi}^{\mathrm{q}}_{\bm{\mathrm{k}}'+\bm{\mathrm{q}}}(\omega'+\nu)\bigl[\psi^{\mathrm{c}}_{\bm{\mathrm{k}}}(\omega)\psi^{\mathrm{c}}_{\bm{\mathrm{k}}'}(\omega')&\nonumber\\[0.5em]
\quad&+\psi^{\mathrm{q}}_{\bm{\mathrm{k}}}(\omega)\psi^{\mathrm{q}}_{\bm{\mathrm{k}}'}(\omega')\bigr]+\mathrm{c.c.}\Bigr)-F_{\mathrm{p}}\left(\bar{\psi}^{\mathrm{q}}_{\bm{0}}(0)+\psi^{\mathrm{q}}_{\bm{0}}(0)\right),&\nonumber 
\end{flalign}
where $D_{\bm{\mathrm{k}}}^0(\omega)=\omega+\omega_{\mathrm{p}}-\epsilon_{\bm{\mathrm{k}}}-\mathrm{i}\kappa$ and $\epsilon_{\bm{\mathrm{k}}}=(\bm{\mathrm{k}}+\bm{\mathrm{k}}_{\mathrm{p}})^2/2m^*$. To
connect this to the normal ordered current response, we add an extra
term to the action, $\delta S$, that contains two source fields,
$\bm{\mathrm{f}}$ and $\bm{\uptheta}$, where the former is coupled to the Keldysh
`quantum' current and the latter to the observable current: 
\begin{flalign}
\delta S[\bm{\mathrm{f}},\bm{\uptheta}]&=\sum_{\omega,\bm{\mathrm{k}},\bm{\mathrm{q}}}\gamma_i(\bm{\mathrm{k}}+\bm{\mathrm{q}},\bm{\mathrm{k}})\bar{\Psi}_{\bm{\mathrm{k}}+\bm{\mathrm{q}}}(\omega)[\hat{\sigma}^{\mathrm{K}}_xf_{i}(\bm{\mathrm{q}})&\\[0.5em]
&+(\hat{\sigma}_z^{\mathrm{K}}+\mathrm{i}\hat{\sigma}_y^{\mathrm{K}})\theta_{i}(\bm{\mathrm{q}})]\Psi_{\bm{\mathrm{k}}}(\omega),&\nonumber
\end{flalign}
where $\sigma_i^{\mathrm{K}}$ are the Pauli matrices in the Keldysh basis.\\

\noindent\textbf{Current-current response function.} Having
constructed a path integral, $\mathcal{Z}=\int\mathcal{D}(\bar{\Psi},\Psi)\e^{(\mathrm{i}S+\mathrm{i}\delta S)}$, the response function is found by
differentiating it with respect to the source fields:
\beq
\chi_{ij}(\bm{\mathrm{q}})=\left.-\frac{\mathrm{i}}{2}\frac{d^2\mathcal{Z}[\bm{\mathrm{f}},\bm{\uptheta}]}{df_{i}(\bm{\mathrm{q}})d\theta_{j}(-\bm{\mathrm{q}})}\right|_{\bm{\mathrm{f}}=\bm{\uptheta}=\bm{0}}.
\eeq
Owing to the interaction term in the Hamiltonian, this calculation requires that we make the substitution $\Psi=\Psi_0+\delta\Psi$ for a mean-field and quadratic fluctuations, modifying our path integral,
\beq
\mathcal{Z}=\int\mathcal{D}(\delta\bar{\Psi},\delta\Psi)\exp[\mathrm{i}S_0+\mathrm{i}\sum\delta\bar{\Psi}(\mathcal{D}^{-1}+A[\bm{\mathrm{f}},\bm{\uptheta}])\delta\Psi],
\eeq
where $\mathcal{D}^{-1}$ is the inverse matrix of Green's
functions and $A[\bm{\mathrm{f}},\bm{\uptheta}]$ consists of the fluctuation
terms dependent on the source fields. In general, our mean-field will be dependent on the source fields and integrating out the
fluctuations we find our response function is given by
\beqy
\chi_{ij}(\bm{\mathrm{q}})&=&\chi_{ij}^{\mathrm{mf}}(\bm{\mathrm{q}})+\chi_{ij}^{\mathrm{fl}}(\bm{\mathrm{q}})\\[0.5em]
&=&-\frac{\mathrm{i}}{2}\bigg[\mathrm{i}\frac{d^2S_0}{df_i(\bm{\mathrm{q}})d\theta_j(-\bm{\mathrm{q}})}\\[0.5em]
&&+\frac{1}{2}\Tr\left(\mathcal{D}\frac{dA}{df_i(\bm{\mathrm{q}})}\mathcal{D}\frac{dA}{d\theta_j(-\bm{\mathrm{q}})}\right)\nonumber\\[0.5em]
&&-\frac{1}{2}\Tr\left(\mathcal{D}\frac{d^2A}{df_i(\bm{\mathrm{q}})d\theta_j(-\bm{\mathrm{q}})}\right)\bigg]\nonumber,
\eeqy
\noindent where the first term comes from the mean-field and the
others from the fluctuations.\\ 

\noindent \textbf{Coefficients.} Using the shorthands, 

\begin{flalign}
J(\bm{\mathrm{q}})&\equiv\omega_{\mathrm{p}}-\epsilon_{\bm{\mathrm{q}}}+\mathrm{i}\kappa-2V\abs{\psi_0}^2,&
\end{flalign}
\begin{flalign}
K(\bm{\mathrm{q}})&\equiv-\frac{2}{\det[(D^{\mathrm{R}})^{-1}(\omega=0,\bm{\mathrm{q}})]}&\\[0.5em]
&=-\frac{2}{J(\bm{\mathrm{q}})J^*(-\bm{\mathrm{q}})-V^2\abs{\psi_0}^4},&
\end{flalign}
\begin{flalign}
L_1&\equiv J(\bm{0})+J^*(\bm{0})+2V\abs{\psi_0}^2=2\Delta-2V\abs{\psi_0}^2,&
\end{flalign}
\begin{flalign}
L_2&\equiv J(\bm{0})+V\abs{\psi_0}^2=\Delta+\mathrm{i}\kappa-V\abs{\psi_0}^2,&
\end{flalign}\smallskip

\noindent where $\epsilon_{\bm{\mathrm{q}}}=(\bm{\mathrm{q}}+\bm{\mathrm{k}}_{\mathrm{p}})^2/2m^*$, the coefficients in Eqs. (7-10) are given by:
\begin{flalign}
c^{\mathrm{mf}}_{+,+}(\bm{\mathrm{q}})&=\frac{1}{2}\abs{\psi_0}^2K(\bm{\mathrm{q}})J^*(-\bm{\mathrm{q}}),&
\end{flalign}
\begin{flalign}
c^{\mathrm{mf}}_{+,-}(\bm{\mathrm{q}})&=c^{(0)}_{-,+}(\bm{\mathrm{q}})=\frac{1}{2}\abs{\psi_0}^2K(\bm{\mathrm{q}})V\abs{\psi_0}^2,&
\end{flalign}
\begin{flalign}
c^{\mathrm{mf}}_{-,-}(\bm{\mathrm{q}})&=\frac{1}{2}\abs{\psi_0}^2K(\bm{\mathrm{q}})J(\bm{\mathrm{q}}),&
\end{flalign}
\begin{widetext}
\begin{flalign}
\mathcal{C}^{\mathcal{A}}_{+}(\bm{\mathrm{q}})&=-VK(\bm{\mathrm{q}})
\bmat
\abs{\psi_0}^2[J^*(-\bm{\mathrm{q}})+V\abs{\psi_0}^2]&\psi_0^2J^*(-\bm{\mathrm{q}})\\
V\bar{\psi}_0^3\psi_0&\abs{\psi_0}^2[J^*(-\bm{\mathrm{q}})+V\abs{\psi_0}^2]
\emat
,&
\end{flalign}
\begin{flalign}
\mathcal{C}^{\mathcal{A}}_{-}(\bm{\mathrm{q}})&=-VK(\bm{\mathrm{q}})
\bmat
\abs{\psi_0}^2[J(\bm{\mathrm{q}})+V\abs{\psi_0}^2]&V\bar{\psi}_0\psi_0^3\\
\bar{\psi}_0^2J(\bm{\mathrm{q}})&\abs{\psi_0}^2[J(\bm{\mathrm{q}})+V\abs{\psi_0}^2]
\emat
,&
\end{flalign}
\begin{flalign}
\mathcal{C}^{\mathcal{B}}_{+}(\bm{\mathrm{q}})&=-VK(\bm{\mathrm{q}})
\bmat
\abs{\psi_0}^2[J^*(-\bm{\mathrm{q}})+V\abs{\psi_0}^2]&V\bar{\psi}_0\psi_0^3\\
\bar{\psi}_0^2J^*(-\bm{\mathrm{q}})&\abs{\psi_0}^2[J^*(-\bm{\mathrm{q}})+V\abs{\psi_0}^2]
\emat
,&
\end{flalign}
\begin{flalign}
\mathcal{C}^{\mathcal{B}}_{-}(\bm{\mathrm{q}})&=-VK(\bm{\mathrm{q}})
\bmat
\abs{\psi_0}^2[J(\bm{\mathrm{q}})+V\abs{\psi_0}^2]&\psi_0^2J(\bm{\mathrm{q}})\\
V\bar{\psi}_0^3\psi_0&\abs{\psi_0}^2[J(\bm{\mathrm{q}})+V\abs{\psi_0}^2]
\emat
,&
\end{flalign}
\begin{flalign}
\mathcal{C}^{\mathcal{M}}_{+,+}(\bm{\mathrm{q}})&=-\frac{1}{2}VK(\bm{\mathrm{q}})
\bmat
\abs{\psi_0}^2\big(K(\bm{\mathrm{q}})[1-K(\bm{0})V\abs{\psi_0}^2L_1]&\psi_0^2\big(-K(\bm{0})K(\bm{\mathrm{q}})V\abs{\psi_0}^2L_2\\
\times[J^*(-\bm{\mathrm{q}})^2+V^2\abs{\psi_0}^4]&\times[J^*(-\bm{\mathrm{q}})^2+V^2\abs{\psi_0}^4]\\
-K(\bm{0})J^*(-\bm{\mathrm{q}})L_1[K(\bm{\mathrm{q}})V^2\abs{\psi_0}^4-1]\big)&+J^*(-\bm{\mathrm{q}})(K(\bm{\mathrm{q}})V\abs{\psi_0}^2\\
&-K(\bm{0})L_2[K(\bm{\mathrm{q}})V^2\abs{\psi_0}^4-1])\big)\\[0.5em]
\bar{\psi}_0^2\big(-K(\bm{0})K(\bm{\mathrm{q}})V\abs{\psi_0}^2L^*_2&\abs{\psi_0}^2\big(K(\bm{\mathrm{q}})[1-K(\bm{0})V\abs{\psi_0}^2L_1]\\
\times[J^*(-\bm{\mathrm{q}})^2+V^2\abs{\psi_0}^4]&\times[J^*(-\bm{\mathrm{q}})^2+V^2\abs{\psi_0}^4]&\\
+J^*(-\bm{\mathrm{q}})(K(\bm{\mathrm{q}})V\abs{\psi_0}^2&-K(\bm{0})J^*(-\bm{\mathrm{q}})L_1[K(\bm{\mathrm{q}})V^2\abs{\psi_0}^4-1]\big)\\
-K(\bm{0})L_2^*[K(\bm{\mathrm{q}})V^2\abs{\psi_0}^4-1])\big)&
\emat
,&
\end{flalign}
\begin{flalign}
\mathcal{C}^{\mathcal{M}}_{+,-}(\bm{\mathrm{q}})&=-\frac{1}{2}VK(\bm{\mathrm{q}})
\bmat
V\abs{\psi_0}^4\big(K(\bm{\mathrm{q}})[1-K(\bm{0})V\abs{\psi_0}^2L_1]&\psi_0^2\big(-K(\bm{0})V\abs{\psi_0}^2(K(\bm{\mathrm{q}})V\abs{\psi_0}^2L_2\\
\times[J(\bm{\mathrm{q}})+J^*(-\bm{\mathrm{q}})]&\times[J(\bm{\mathrm{q}})+J^*(-\bm{\mathrm{q}})]+K(\bm{\mathrm{q}})J(\bm{0})J(\bm{\mathrm{q}})J^*(-\bm{\mathrm{q}})\\
-K(\bm{0})([K(\bm{\mathrm{q}})V^2\abs{\psi_0}^4-2]L_2^*&+K(\bm{\mathrm{q}})V^3\abs{\psi_0}^6-2V\abs{\psi_0}^2)\\
+K(\bm{\mathrm{q}})J(\bm{\mathrm{q}})J^*(-\bm{\mathrm{q}})L_2&+K(\bm{\mathrm{q}})J(\bm{\mathrm{q}})J^*(-\bm{\mathrm{q}})\big)\\[0.5em]
\bar{\psi}_0^2\big(-K(\bm{0})V\abs{\psi_0}^2(K(\bm{\mathrm{q}})V\abs{\psi_0}^2L_2^*&V\abs{\psi_0}^4\big(K(\bm{\mathrm{q}})[1-K(\bm{0})V\abs{\psi_0}^2L_1]\\
\times[J(\bm{\mathrm{q}})+J^*(-\bm{\mathrm{q}})]+K(\bm{\mathrm{q}})J^*(\bm{0})V^2\abs{\psi_0}^4&\times[J(\bm{\mathrm{q}})+J^*(-\bm{\mathrm{q}})]\\
+K(\bm{\mathrm{q}})J(\bm{\mathrm{q}})J^*(-\bm{\mathrm{q}})V\abs{\psi_0}^2-2J^*(\bm{0}))&-K(\bm{0})([K(\bm{\mathrm{q}})V^2\abs{\psi_0}^4-2]L_2^*\\
+K(\bm{\mathrm{q}})V^2\abs{\psi_0}^4\big)&+K(\bm{\mathrm{q}})J(\bm{\mathrm{q}})J^*(-\bm{\mathrm{q}})L_2
\emat
,&
\end{flalign}
\begin{flalign}
\mathcal{C}^{\mathcal{M}}_{-,+}(\bm{\mathrm{q}})&=-\frac{1}{2}VK(\bm{\mathrm{q}})
\bmat
V\abs{\psi_0}^4\big(K(\bm{\mathrm{q}})[1-K(\bm{0})V\abs{\psi_0}^2L_1]&\psi_0^2\big(-K(\bm{0})V\abs{\psi_0}^2(K(\bm{\mathrm{q}})V\abs{\psi_0}^2L_2\\
\times[J(\bm{\mathrm{q}})+J^*(-\bm{\mathrm{q}})]&\times[J(\bm{\mathrm{q}})+J^*(-\bm{\mathrm{q}})]+K(\bm{\mathrm{q}})J(\bm{0})V^2\abs{\psi_0}^4\\
-K(\bm{0})([K(\bm{\mathrm{q}})V^2\abs{\psi_0}^4-2]L_2&+K(\bm{\mathrm{q}})J(\bm{\mathrm{q}})J^*(-\bm{\mathrm{q}})V\abs{\psi_0}^2-2J(\bm{0}))\\
+K(\bm{\mathrm{q}})J(\bm{\mathrm{q}})J^*(-\bm{\mathrm{q}})L_2^*&+K(\bm{\mathrm{q}})V^2\abs{\psi_0}^4\big)\\[0.5em]
\bar{\psi}_0^2\big(-K(\bm{0})V\abs{\psi_0}^2(K(\bm{\mathrm{q}})V\abs{\psi_0}^2L_2^*&V\abs{\psi_0}^4\big(K(\bm{\mathrm{q}})[1-K(\bm{0})V\abs{\psi_0}^2L_1]\\
\times[J(\bm{\mathrm{q}})+J^*(-\bm{\mathrm{q}})]+K(\bm{\mathrm{q}})J^*(\bm{0})J(\bm{\mathrm{q}})J^*(-\bm{\mathrm{q}})&\times[J(\bm{\mathrm{q}})+J^*(-\bm{\mathrm{q}})]\\
+K(\bm{\mathrm{q}})V^3\abs{\psi_0}^6-2V\abs{\psi_0}^2)&-K(\bm{0})([K(\bm{\mathrm{q}})V^2\abs{\psi_0}^4-2]L_2\\
+K(\bm{\mathrm{q}})J(\bm{\mathrm{q}})J^*(-\bm{\mathrm{q}})\big)&+K(\bm{\mathrm{q}})J(\bm{\mathrm{q}})J^*(-\bm{\mathrm{q}})L_2^*
\emat
,&
\end{flalign}
\begin{flalign}
\mathcal{C}^{\mathcal{M}}_{-,-}(\bm{\mathrm{q}})&=-\frac{1}{2}VK(\bm{\mathrm{q}})
\bmat
\abs{\psi_0}^2\big(K(\bm{\mathrm{q}})[1-K(\bm{0})V\abs{\psi_0}^2L_1]&\psi_0^2\big(-K(\bm{0})K(\bm{\mathrm{q}})V\abs{\psi_0}^2L_2\\
\times[J(\bm{\mathrm{q}})^2+V^2\abs{\psi_0}^4]&\times[J(\bm{\mathrm{q}})^2+V^2\abs{\psi_0}^4]\\
-K(\bm{0})J(\bm{\mathrm{q}})L_1[K(\bm{\mathrm{q}})V^2\abs{\psi_0}^4-1]\big)&+J(\bm{\mathrm{q}})(K(\bm{\mathrm{q}})V\abs{\psi_0}^2\\
&-K(\bm{0})L_2[K(\bm{\mathrm{q}})V^2\abs{\psi_0}^4-1])\big)\\[0.5em]
\bar{\psi}_0^2\big(-K(\bm{0})K(\bm{\mathrm{q}})V\abs{\psi_0}^2L^*_2&\abs{\psi_0}^2\big(K(\bm{\mathrm{q}})[1-K(\bm{0})V\abs{\psi_0}^2L_1]\\
\times[J(\bm{\mathrm{q}})^2+V^2\abs{\psi_0}^4]&\times[J(\bm{\mathrm{q}})^2+V^2\abs{\psi_0}^4]&\\
+J(\bm{\mathrm{q}})(K(\bm{\mathrm{q}})V\abs{\psi_0}^2&-K(\bm{0})J(\bm{\mathrm{q}})L_1[K(\bm{\mathrm{q}})V^2\abs{\psi_0}^4-1]\big)\\
-K(\bm{0})L_2^*[K(\bm{\mathrm{q}})V^2\abs{\psi_0}^4-1])\big)&
\emat
.&
\end{flalign}
\end{widetext}
}

\noindent\textbf{\large Acknowledgements \normalsize}\\ 
We would like to thank I. Carusotto and K. Dunnett for helpful discussions, and A. Zamora for creating Fig.~\ref{fig:polaritons}(a). MHS acknowledges financial support from EPSRC (Grants No. EP/I028900/2 and No. EP/K003623/2) and  JK from EPSRC program “Hybrid Polaritonics” (EP/M025330/1).

\end{document}